\documentclass[twocolumn,showpacs,aps,prl,superscriptaddress,unsortedaddress]{revtex4}
\usepackage{graphicx}
\usepackage{epsf}
\newcommand{\etal}{{\it et al.}}

\begin{document}

\title{Evidence for pairing above $T_c$ from the dispersion in the pseudogap phase of cuprates}

\author{A. Kanigel}
\affiliation{Department of Physics, University of Illinois at Chicago, Chicago, IL 60607}
\affiliation{Department of Physics, Technion, Haifa 32000, Israel}
\author{U. Chatterjee}
\affiliation{Department of Physics, University of Illinois at Chicago, Chicago, IL 60607}
\author{M. Randeria}
\affiliation{Department of Physics, The Ohio State University, Columbus, OH  43210}
\author{M. R. Norman}
\affiliation{Materials Science Division, Argonne National Laboratory, Argonne, IL 60439}
\author{G. Koren}
\affiliation{Department of Physics, Technion, Haifa 32000, Israel}
\author{K. Kadowaki}
\affiliation{Institute of Materials Science, University of Tsukuba, Ibaraki 305-3573, Japan}
\author{J. C. Campuzano}
\affiliation{Department of Physics, University of Illinois at Chicago, Chicago, IL 60607}
\affiliation{Materials Science Division, Argonne National Laboratory, Argonne, IL 60439}

\begin{abstract}

In the underdoped high temperature superconductors, instead of a complete Fermi surface above $T_c$, only disconnected Fermi arcs appear, separated by regions that still exhibit an energy
gap. We show that in this pseudogap phase, the energy-momentum relation of electronic excitations near $E_F$ behaves like the dispersion of a normal metal on the Fermi arcs, but like that of a superconductor in the gapped regions. We argue that this dichotomy in the dispersion is hard to reconcile with a competing order parameter, but is consistent with pairing without condensation.
\end{abstract}

\date{\today}
\maketitle

\begin{figure}
\begin{center}
\includegraphics[width=8.5cm]{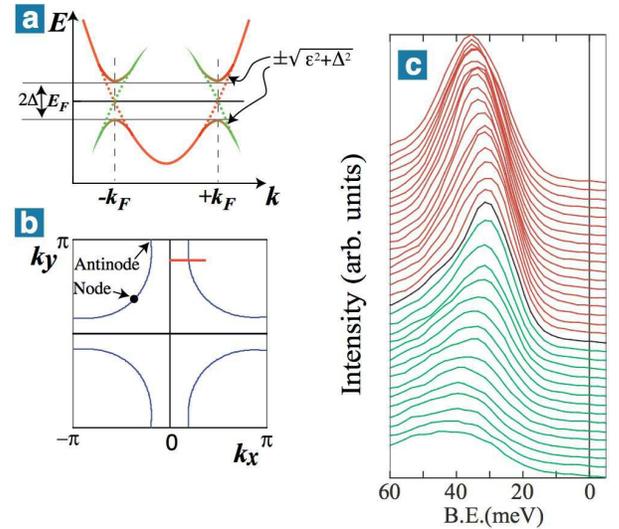}
\end{center}
\caption{
(Color online) a) Schematic diagram of the normal state dispersion (dashed curve) which acquires a gap with a characteristic Bogoliubov dispersion in the superconducting state (solid curves). The Bogoliubov dispersion arises from particle-hole mixing, which is the mixing of electron
and hole states.
b) The Fermi surface in the Brillouin zone, identifying where the d-wave gap is zero (node) and where it is maximal (anti-node). c) Energy distribution curves (EDCs) in the superconducting state ($T$ = 17 K) of a thin film of Bi$_2$Sr$_2$CaCu$_2$O$_8$ (Bi2212) sample with $T_c$ = 80 K along the momentum cut identified in (b). Each curve corresponds to an increase in momentum of 0.003 \AA$^{-1}$. The EDC at ${\bf k}_F$ is indicated by the thick curve.
}
\label{Fig1}
\end{figure}

There is no consensus regarding the origin of the pseudogap \cite{timusk,ding96,loeser96} in underdoped cuprates. The arguments can be distilled into two general ideas \cite{adv-phys}: 
the pseudogap arises either from pairing of electrons \cite{randeria,emery}
in a state precursor to superconductivity, or from an alternate order parameter \cite{DDW,Varma, Lee}. 
Lacking a direct measurement of the momentum-dependent pairing correlations, we ask whether some features unique to Cooper paring are present in the electronic excitation spectrum above $T_c$. In particular, the dispersion of states of energy $\epsilon$ and momentum ${\bf k}$ in the normal state is usually linear for a small energy interval near the Fermi energy, shown schematically in Fig.~\ref{Fig1}a as a parabola.

The locus of Fermi crossings, the Fermi surface, is shown in Fig.~\ref{Fig1}b.
In the superconducting state, the linear dispersion transforms into the Bogoliubov dispersion $E_k=\pm \sqrt{\epsilon_k^2+\Delta_k^2}$, shown as solid curves in Fig.~\ref{Fig1}a, where $\Delta_{\bf k}$ is the gap function.  The minimum in the excitation energy along a momentum cut normal to the Fermi surface is at $|\Delta_{\bf k}|$, which occurs at ${\bf k}_F$, the Fermi momentum of the normal state.  This is a consequence of the fact that the pairs condense with a zero center of mass momentum.  

The Bogoliubov dispersion below $T_c$ can be readily observed in experimental angle resolved photoemission (ARPES) spectra \cite{PH}, as shown in Fig.~\ref{Fig1}c. The excitation energy, $E_{\bf k}$, approaches $E_F$, but instead of crossing it, it reaches a minimum value at ${\bf k}_F$, before receding away from $E_F$, where it remains only visible for a small range of ${\bf k}$ beyond ${\bf k}_F$. These new states  result from the mixing of electrons
with holes,
as shown in Fig.~\ref{Fig1}a. Even the Bogoliubov dispersion branch above
$E_F$ has been seen in ARPES by thermal population \cite{Matsui}.

\begin{figure}
\begin{center}
\includegraphics[width=8.5cm]{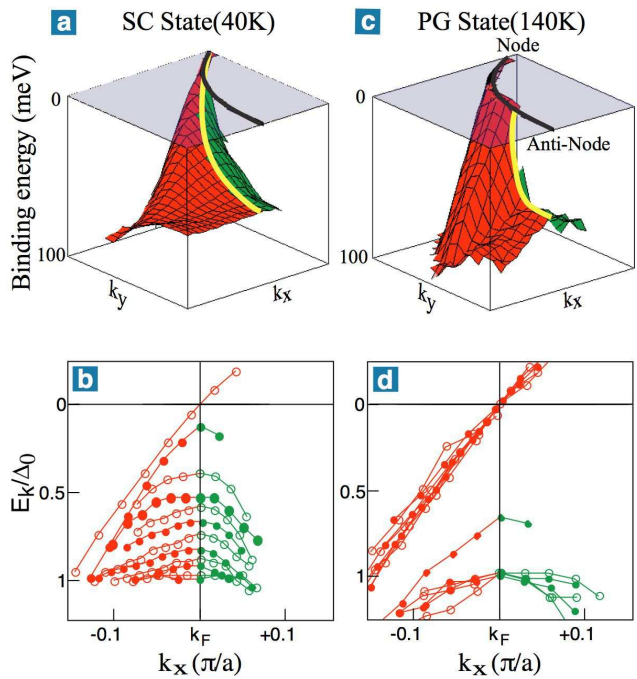}
\end{center}
\caption{
(Color online) (a) Binding energy of the spectral peak from Fermi function divided ARPES data for a  Bi2212 single crystal with $T_c=90$ K as a function of ($k_x$,$k_y$) in the superconducting state at $T$=40 K; note that the peak crosses $E_F$ (the top face of the cube) only at the nodal point. In the rest of the zone the peak reaches a finite minimum in binding energy, and then recedes to higher energy, losing intensity in the process.
The dispersion minima are plotted as a
curve on the dispersion surface, with its projection onto the
top face also shown, which is equivalent to the normal state Fermi surface. (b) Cuts for various $k_y$ from (a), showing the peak dispersion versus $k_x$.
c)  Data same as in (a), but in the pseudogap phase ($T$=140 K).  Here the spectral peak crosses $E_F$ for an extended length, creating a Fermi arc. Surprisingly, around the anti-node, where the spectral gap survives, it can be clearly seen that the dispersion bends back as in the superconducting state. (d)  Cuts for various $k_y$ from (c), showing the peak dispersion vs $k_x$. Note that the dispersion either crosses $E_F$ (along the Fermi arc), or exhibits Bogoliubov-like behavior similar to the superconducting state.
}
\label{Fig2}
\end{figure}
 
An important consequence of particle-hole mixing with zero center of mass momentum is that the minimum gap location is identical to the normal state Fermi momentum. Fig.~\ref{Fig2}a shows the dispersion in the superconducting state at $T=40$ K for a $T_c$ = 90 K sample over the entire Brillouin zone.  This dispersion was obtained as follows. The ARPES spectrum is proportional to the product of the single-particle spectral function and the Fermi function, convolved with the instrumental resolution. We therefore take the raw spectra, and divide this by a resolution broadened Fermi function obtained by  fitting a reference gold spectrum in electrical contact with the sample that is used to determine the chemical potential.  The peak positions of these divided spectra define our dispersion, which we show below $E_F$ in Fig.~\ref{Fig2}a. The
forward surface in this plot represents the particle-like region of the dispersion, while the hole-like region is behind.
The  top face of the cube is the chemical potential ($E_F$). The spectral peak crosses $E_F$ only at one point (the node), while the spectra at other momenta are gapped. The
curve on the dispersion surface shows the dispersion minima, which to a good approximation follow the simple d-wave gap function $|\Delta_{\bf k}|=\Delta_0|\cos{k_xa}-\cos {k_ya}|/2$ where $\Delta_0$ is the maximum d-wave gap. The
curve on the top face shows the location where these minima occur in the Brillouin zone, which coincides with the normal state Fermi surface, ${\bf k}_F$. In Fig.~\ref{Fig2}b we show cuts from Fig.~\ref{Fig2}a at regular $k_y$ intervals, where the back bending is clearly seen in all cuts for $k_x$ beyond $k_F$, except at the node. 

We now turn to the pseudogap phase, where in Fig.~\ref{Fig2}c we show the dispersion taken at $T$=140 K. In contrast to the superconducting state, the dispersion now crosses $E_F$ for an extended length, forming a Fermi arc \cite{Nat98,Kanigel}. This arc extends from the node to approximately half way to the anti-node.  The rest of the spectra are gapped. Moreover, we find the remarkable fact that where it is gapped, the dispersion shows back bending characteristic of the superconducting state.
In Fig.~\ref{Fig2}d,  the momentum cuts of the dispersion for various $k_y$ in the pseudogap phase are shown. Note, that for each cut where a gap exists, the bending back behavior is present. 

\begin{figure}
\includegraphics[width=8.5cm]{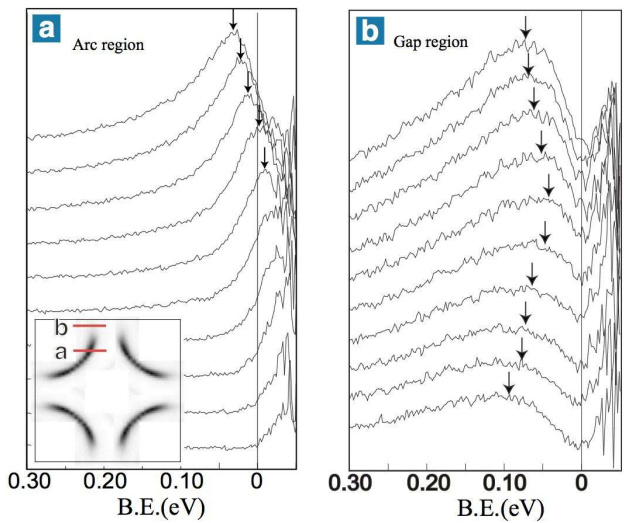}
\caption{
(Color online) Dispersion of the Fermi function divided EDCs for a Bi2212 single crystal with $T_c$=90 K in the pseudogap phase ($T$=140 K) along (a) cut `a' shown in the inset which crosses the Fermi arc (the gray shading in the inset indicates the ARPES intensity at $E_F$), and (b) cut `b' shown in the inset of (a) that crosses $k_F$ in the gapped region.
}
\label{Fig3}
\end{figure}

This is a remarkable situation - the dispersion in part of the Brillouin zone (on the arc) behaves as if the sample were a normal metal, while in the remainder of the zone, the dispersion behaves as if the sample were superconducting, even though we are above $T_c$. To further emphasize this dichotomy, we show in Fig.~\ref{Fig3}a the Fermi function divided spectra for a  momentum cut  through the Fermi arc, as indicated in the inset. The spectral peak disperses through $E_F$ in the normal state, showing no indication of a gap. The relatively high sample temperature of 140 K allows us to follow the dispersion for some distance above $E_F$. The spectral peak is never observed to bend back. In contrast, in a cut through the gapped portion of the normal state Fermi surface shown in Fig.~\ref{Fig3}b, the dispersion (arrows) clearly exhibits the characteristic bending back of the superconducting
state.  The spectral peak approaches $E_F$, reaches a minimum, and then recedes.  

\begin{figure}
\includegraphics[width=7cm]{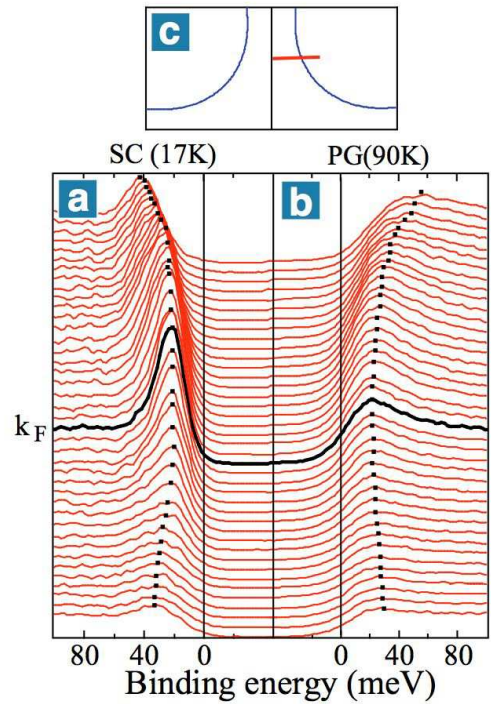}
\caption{(Color online)
EDCs in (a) the superconducting state ($T$=17 K) and (b) the pseudogap phase ($T$=90 K) for the cut in the zone shown in (c).  The marked curves in (a) and (b) represent the minimum gap location, and can
 be seen to occur at the same value of $k$, corresponding to the normal state $k_F$.  Note that for this cut, the Fermi surface has already started to curve away from
the flat regions near the anti-node.
}
\label{Fig4}
\end{figure}

It is important to compare the dispersions in the superconducting and pseudogap phases. In Fig.~\ref{Fig4}a,b we show raw photoemission spectra (not divided by the Fermi function) from a Bi2212 film with $T_c =$ 80K, taken at the momentum cut shown in Fig.~\ref{Fig4}c. The superconducting state spectra are shown to the left of zero energy, while the pseudogap phase spectra are reflected to the right of zero energy. Each pair of curves in the superconducting and pseudogap phases were obtained at the same ${\bf k}$ point. A key difference between panels (a) and (b) is in the spectral linewidths: there is a narrow quasiparticle peak below $T_c$, while the pseudogap spectra are broad, indicating a short electronic lifetime [14].   Nevertheless, the dispersions in both panels exhibit the same characteristic bending back behavior. More importantly, the dispersion minima in both the pseudogap and superconducting phases occur at the \textbf{same} ${\bf k}$,
which happen to be the ${\bf k}_F$ of the normal state \cite{PH}. We find that for all samples we have investigated, the minimum gap occurs exactly at the same ${\bf k}$ above and below $T_c$ for all cuts, irrespective of their position along the Fermi surface.  

\begin{figure}
\includegraphics[width=7.5cm]{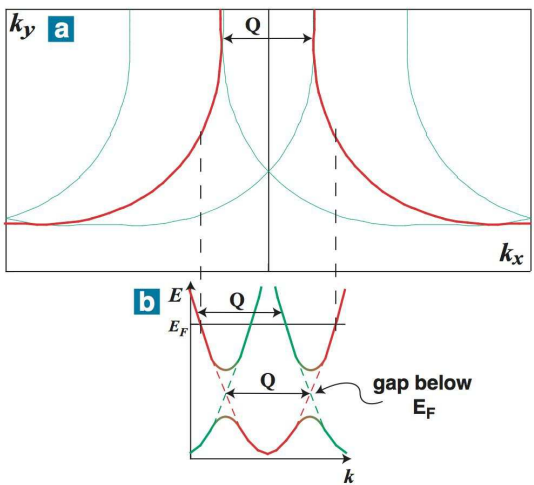}
\caption{
(Color online)
a) The Fermi surface
curves umklapped by $\pm {\bf Q}$,
with ${\bf Q}$ chosen so as to span the Fermi surface in its flat parts near the anti-node. (b) Schematic diagram of the dispersion along a cut where the separation between the two Fermi surfaces becomes larger than $|Q|$, that is between where the dashed vertical lines intersect the Fermi surface in (a).  
Note that a gap forms below the Fermi energy, and thus would be observable by ARPES if it existed.
}
\label{Fig5}
\end{figure}

We now discuss the implications of our results. It has been suggested that the pseudogap originates from some ordering phenomenon  -- unrelated to superconductivity -- characterized by a wavevector ${\bf Q}$. Let us look at the case \cite{KShen} of a ${\bf Q}$ that spans flat parts of the Fermi surface near the anti-node, where the pseudogap is maximal, as shown in Fig.~\ref{Fig5}a. This, and other suggestions regarding the origin of the pseudogap, requires us to consider whether evidence for such a ${\bf Q}$ vector exists in the data. As soon as the spanning vector between the Fermi surfaces becomes significantly longer than ${\bf Q}$, the states near $E_F$ will no longer be gapped. However, as shown in Fig.~\ref{Fig5}b, the dispersion for
a momentum cut where the Fermi surfaces are separated by a vector longer than ${\bf Q}$ (i.e., for a k-point not on the flat part of the Fermi surface) still shows back bending after reaching a minimum at the same ${\bf k}$ as the normal state ${\bf k}_F$. 

Moreover, the alteration of the dispersion resulting from an ordering vector ${\bf Q}$ is not limited to the  region near ${\bf k}_F$ and $E_F$, as it is in the case of superconductivity. Because ${\bf Q}$ mixes states at ${\bf k}$ with those at ${\bf k}\pm {\bf Q}$, additional Fermi sheets are present which are images of the Fermi surface displaced by $\pm {\bf Q}$.
We do not see these `umklapps', even close to the Fermi surface
where the mixing is strong enough that their intensity should be observable by ARPES, as they are in metals which do exhibit a density wave instability \cite{Grioni}.  Moreover, such ordering has the unavoidable consequence of shifting the gap \textit{away} from $E_F$ when the Fermi surfaces are spanned by a vector longer than ${\bf Q}$, as shown in Fig.~\ref{Fig5}b. Since the Fermi surface is hole-like, this gap appears \textit{below} $E_F$ and thus should be observed by ARPES if it existed.  We find no evidence for such a gap.

We emphasize that these arguments eliminate \emph{any} ${\bf Q} \ne 0$ order,
including ${\bf Q} = (\pi,\pi)$, and not just the specific ${\bf Q}$ of Fig.~\ref{Fig5},
as the origin of the pseudogap. The observed bending back of the dispersion is
also inconsistent with that predicted for a ${\bf Q}$=0 order parameter due to orbital currents \cite{Varma}. We are thus left with the only remaining possibility that the observed Bogoliubov-like dispersion above $T_c$  is the analog of particle-hole mixing below $T_c$ arising due to short-range superconducting order. This would naturally explain why the minimum gap \textit{always} occurs at ${\bf k}_F$ in momentum space and why the corresponding spectral function has a minimum at $E_F$. 

In conventional BCS superconductors, the breaking up of the Cooper pairs is responsible for the phase transition at $T_c$, since the energy gap is much smaller than the phase stiffness, which is controlled by the superfluid density. In contrast, as suggested early on by Uemura {\etal}~\cite{Uemura}, it is the small superfluid density in the underdoped cuprates which determines the loss of phase coherence at $T_c$, an idea which is further substantiated by recent measurements in highly underdoped materials \cite{Broun, Hetel}. In these materials, the pairing gap is much larger than the superfluid density, and thus pairing survives above $T_c$. Recent examples of possible observations of pairing without phase coherence are in systems as diverse as granular superconductors \cite{Markovic} and cold atomic Fermi gases \cite{Schunck}.

To what extent are other experiments on underdoped cuprates consistent with
the idea of pairing of electrons above $T_c$ in the pseudogap phase? Early NMR experiments \cite{NMR1,NMR2} showed a
freezing out of the spin susceptibility and the relaxation rate $1/T_1 T$ with decreasing $T$.
This directly implies the formation of singlet pairs with an onset temperature well above
$T_c$, and consistent with $T^\ast$ at which the pseudogap becomes observable in ARPES \cite{ding96,loeser96}
and in the STM tunneling density of states \cite{STM}.
More recently there have been two important experiments on the existence of fluctuating superconducting
regions in the pseudogap phase: the direct observation of diamagnetism above $T_c$ \cite{Ong1}
and an anomalously large Nernst signal \cite{Ong2} attributed to vortices above $T_c$.
The Nernst onset temperature, though larger than $T_c$, is definitely lower than $T^\ast$ and has a different doping dependence;
it goes to zero close to the doping where superconductivity disappears, while $T^*$ and the ARPES pseudogap 
continue to increase in magnitude with underdoping. We believe that there is no contradiction here; in addition to pairing,
the Nernst effect also needs local phase coherence over large enough spatial regions for the vortices to exist.

We note that at the present time there is no complete theory of the remarkable dichotomy of the dispersion that we observe in different parts of momentum space, including the temperature dependence of the arcs \cite{Kanigel} and the closing of the gap along the arcs versus its filling in elsewhere \cite{Norman_PRB98}. On quite general grounds, we expect that the ARPES spectral function only involves the average $<|\Delta_{\bf k}|^2>$, which is finite even when the phase of the order parameter is fluctuating, thus leading to a characteristic back bending of the dispersion above $T_c$. The {\bf k}-space anisotropy of the pseudogap and dispersion are nevertheless closely linked to the d-wave anisotropy of the gap -- it is expected that the gap around the node will be more susceptible to fluctuations than the gap around the anti-node because of its smaller magnitude. 

In summary, we found a Bogoliubov-like dispersion in the pseudogap phase of the high temperature cuprate superconductors, despite the fact that there are no sharply defined quasiparticles above $T_c$. This anomalous dispersion leads us to conclude that pairing, without long range phase order, underlies the pseudogap below $T^*$.  On the other hand, superconductivity below $T_c$ arises from the locking of the phase of all the pairs forming a condensate with long range order.

This work was supported by NSF DMR-0606255 and
the U.S. DOE, Office of Science, under Contract No.~DE-AC02-06CH11357.
The Synchrotron Radiation Center is supported by NSF DMR-0084402.

\end{document}